\documentclass[a4paper,fleqn]{cas-dc}
\usepackage[numbers]{natbib}
\usepackage{amsfonts}
\usepackage{amsmath}
\usepackage{mathrsfs}
\usepackage{amssymb}
\usepackage{graphicx}
\usepackage{lipsum} 
\usepackage[T1]{fontenc}
\usepackage{epstopdf}
\usepackage{bm}
\usepackage{color}

\def\tsc#1{\csdef{#1}{\textsc{\lowercase{#1}}\xspace}}
\tsc{WGM}
\tsc{QE}
\tsc{EP}
\tsc{PMS}
\tsc{BEC}
\tsc{DE}
\newcommand{\bPsi}{{\bm \Psi}}
\newcommand{\bpsi}{{\bm \psi}}
\newcommand{\bvarphi}{{\bm \varphi}}

\newcommand{\bw}{{\bm w}}
\newcommand{\br}{{\bm r}}
\newcommand{\bk}{{\bm k}}

\newcommand{\halpha}{\hat{\alpha}}

\newcommand{\rev}[1]{\textcolor{black}{{#1}}}

\begin{document}

\let\WriteBookmarks\relax
\def\floatpagepagefraction{1}
\def\textpagefraction{.001}
\shorttitle{Half-vortex solitons}
\shortauthors{C. Wang et~al.}

\title[mode = title]{Half-vortex soliton lattices in spin-orbit-coupled Bose-Einstein condensates with a \rev{quasi}-flat band.}  

\author[1,2]{Chenhui Wang}

\credit{Conceptualization, Methodology, Software, Data analysis, Writing - Original draft preparation}

\author[1]{Yongping Zhang}
\ead{yongping11@t.shu.edu.cn}
\credit{Conceptualization, Software, Methodology,  Writing - Original draft preparation, Supervision}

\author[2,3]{Vladimir V. Konotop}
\ead{vvkonotop@ciencias.ulisboa.pt}

\credit{Conceptualization, Methodology, Writing - Original draft preparation, Supervision}

\affiliation[1]{organization={Institute for Quantum Science and Technology, Department of Physics,  Shanghai University},
                postcode={200444},  
                city={Shanghai},
                country={China}}
 \affiliation[2]{organization={Centro de F\'{i}sica Te\'orica e Computacional, Faculdade de Ci\^encias, Universidade de Lisboa},
 addressline={Campo Grande, Ed. C8},
                postcode={1749-016},  
                city={Lisboa},
                country={Portugal}}
 \affiliation[3]{organization={Departamento de F\'{i}sica, Faculdade de Ci\^encias, Universidade de Lisboa},
 addressline={Campo Grande, Ed. C8},
                postcode={1749-016},  
                city={Lisboa},
                country={Portugal}}
\cortext[cor1]{Corresponding authors}

\begin{abstract}
Periodic potentials with flat bands in their spectra support strongly localized nonlinear excitations. Although a perfectly flat band cannot exist in a continuous systems, a spin–orbit–coupled Bose–Einstein condensate loaded in a Zeeman lattice can realize the \rev{quasi}-flat lowest band with an extremely narrow bandwidth. In such a quasi-flat band, half-vortex solitons become confined within a single lattice cell, enabling the formation of arrays of coupled half-vortex solitons arranged of various spatial geometries. In this work, we study the existence and stability of these lattices within the framework of the two-component Gross–Pitaevskii equation. We demonstrate that, near the \rev{quasi-}flat band, half-vortex solitons and their arrays can be excited with a nearly negligible number of atoms and are constrained by their local symmetries, which are isomorphic to a dihedral group
of order 8. This allows observation of the respective field patterns in the nearly linear regime where they exhibit enhanced stability. The constructed lattices may have diverse geometric profiles, and in particular create a composite super-half-vortex soliton with nonlinear symmetry breaking.
\end{abstract}

 \begin{highlights}
 \item A two-dimensional Zeeman lattice with a quasi-flat is obtained
 \item Half-vortex solitons with a negligible threshold energy are found; they can be observed in a quasi-linear regime 
\item Stable half-vortex soliton lattices of different shapes, as well as super-half-vortex solitons are obtained
\end{highlights}


\maketitle

\section{Introduction}
Vortex solitons are fundamental two-dimensional (2D) spatially localized excitations carrying embedded vorticity. They have attracted considerable attention over the past few decades (see, e.g., the recent reviews~\cite{Malomed2024,Mihalache2024} and references therein). In particular, it is known that vortex solitons of diverse configurations (i.e., with different component profiles) can be excited in two-component systems such as Bose–Einstein condensates (BECs)~\cite{Han2012,Zamora2012,Kasamatsu2013,Wen2013,Zhang2014} and optical systems with Kerr nonlinearities~\cite{Xu2005,Wang2008}. These systems also support \rev{vortex droplets~\cite{Sanjay2025,Deekshita2025}}, or 2D bright–vortex solitons~\cite{Law2010,Stockhofe2011,Pola2012} characterized by the localization of one component.

Among physical systems supporting localized states with embedded vorticity, spin–orbit coupled (SOC) BECs~\cite{Lin2011, Wu2016, Sun2018} occupy a special place, as they can sustain a specific type of localized excitations, known as half-vortex (or semi-vortex) solitons. Similarly to bright–vortex solitons~\cite{Law2010,Stockhofe2011,Pola2012}, the vorticity in such nonlinear modes is embedded in only one component, while the other component remains free of topological defects. Nevertheless, the two components are coherently coupled and both remain spatially localized. Half-vortex solitons have recently attracted considerable attention~\cite{Kartashov2019,Malomed2019}. In particular, they have been explored in spin–orbit-coupled (SOC) Bose–Einstein condensates (BECs) in free space~\cite{Sakaguchi2014,Sakaguchi2016}, in condensates loaded in Zeeman lattices (ZLs)~\cite{Lobanov2014}, in dipolar systems~\cite{Liao2017,Liu2018}, in BECs with helicoidal SOC~\cite{Kartashov2020a},   and in optical lattices featuring flat bands~\cite{Wang2024a}.

In general, the excitation of localized states in two-dimensional continuous systems (in the absence of structural defects) is characterized by a threshold energy, below which stationary localized solutions do not exist~\cite{Berge1998,Fibich2015}. Exceptions include aperiodic systems, like moir\'e lattices~\cite{Fu2020,Ivanov2023}, quasicrystals~\cite{Wang2024b}, and 2D BECs in low-dimensional 1D SOC~\cite{Kartashov2020} where excitation of solitons is thresholdless. On the other hand, excitation of vortex states is thresholdless~\cite{Vicencio2013} in discrete periodic lattices, when an exact flat band is present in the spectrum. In continuous models, ideal flatness of the bands is only an approximation, as it cannot be achieved exactly~\cite{Leykam2018,Li2025}. However, when a band is quasi-flat (i.e., characterized by an exceptionally small bandwidth), nonlinear modes do not bifurcate from the linear spectrum, yet their excitation threshold can become anomalously low~\cite{Wang2024a,Shen2024}, allowing one to regard them as quasi-linear excitations.
 
The existence of a quasi-flat band (below simply a flat band, for the sake of brevity) is a characteristic feature of SOC-BECs with appropriately chosen parameters~\cite{Zhang2013b,Kartashov2016a,Hui2017}, rather than a consequence of an extreme depth of a lattice that usually justifies the tight-binding approximation. Two-component solitons in such systems in the quasi-linear limit are well approximated by the Wannier functions (WFs) of the respective flat band and have amplitudes linearly dependent on the detuning of the chemical potential towards the respective gap. Since a flat band suppresses the dispersion, a weak nonlinearity is sufficient for sustaining well localized 1D Wannier solitons (WSs)~\cite{Wang2023} and 2D half-vortex solitons~\cite{Wang2024a}. Furthermore, such solitons are exhibit increased stability in the small-amplitude limit.

Strong localization and enhanced stability of half-vortex solitons allow one to consider regular arrays of such solutions (multi-hump solitons) what in 1D case was shown in~\cite{Wang2023} and for a single-component Kerr solitons in 2D superhonycomb lattices featuring a flat band~\cite{Shen2024}.

In this work we report on possibility of creation of stable arrays (lattices) of half-vortex solitons enabled by a flat band of a SOC-BEC
as well as creation of super-half-vortexes, when single half-vortexes are used as building blocks, by analogy with flat-band arrays of optical solitons~\cite{Shen2024,Tai2024}.  
 
The paper is organized as follows. In Sec.~\ref{model}, we formulate the model describing a continuous 2D SOC-BEC loaded in a ZL, which features a quasi-flat band separated by a finite gap from the rest of the spectrum.  In Sec.~\ref{single} we describe families of single half-vortex soliton solutions and perform the stability analysis. In Sec.~\ref{lattice1}, we generate numerically and systematically analyze the half-vortex lattices with an arbitrary size. In Sec.~\ref{super}, stable super–half-vortex solitons are generated numerically. The outcomes are summarized in the conclusion.

\section{The model}
\label{model}

We consider a 2D SOC-BECs which, in the mean-field approximation, is described by the dimensionless Gross-Pitaevskii (GP) equation for the spinor ${\bPsi}=\left({\Psi}_{1},{\Psi}_{2}\right)^{T}$: 
\begin{align}
	\label{GPE}
i\partial_t\bPsi  =H\bPsi+g\left(\bPsi^\dagger\bPsi\right)\bPsi,
\end{align}
Here  
\begin{align}
\label{H}
H=-\frac{1}{2}(\partial_x^2+\partial_y^2)
-i\gamma\left(\sigma_{y}\partial_x -\sigma_{x}\partial_y\right)+\Omega(\br)\sigma_{z},
\end{align}
is the linear Hamiltonian, $g=+1$ ($g=-1$) corresponds to the negative (positive) scattering lengths, $\gamma$ is the strength of the Rashba SOC~\cite{Bychkov1984}, and $\sigma_{x,y,z}$ are the Pauli matrices. The spatially periodic Rabi frequency is given by
\begin{align}
 \label{ZL}
\Omega(\br)=\Omega_0-\Omega_{1}\left[\cos(2x)+\cos(2y)\right],
\end{align}  
where the positive parameters $\Omega_0$ and $\Omega_1$ describe constant Rabi frequency and amplitude of the Zeeman lattice, respectively.
 
In the dimensionless GPE (\ref{GPE}), the spatial coordinate is measured in units of $1/k_L$ where $k_L$ is the projection  of wave vector of the laser beams creating periodic potential onto each of the orthogonal directions (we consider a square lattice). Respectively, $\Omega_{0,1}$ are measured in the units of the recoil frequency $\omega_R=2E_L/\hbar$, where $E_L=\hbar^{2}k_{L}^{2}/2m$ is the  recoil energy, and $m$ is the atomic mass. The strength of effective SOC $\gamma$  can be varied by adjusting the Raman beams inducing coupling \cite{Jimenez2015,Zhang2013a}. The physical number of condensate atoms $\cal N$ is determined by the norm $N=\int\bPsi^\dagger\bPsi d\br$: ${\cal N}= N N_0$, where $N_0=E_L\sqrt{2\pi\hbar\omega_z/m}/(\hbar k_L^2|a_s|)$, $\omega_{z}$ is the transverse trap frequency, and $a_{s}$ is s-wave scattering length. The orders of magnitude of the dimensionless quantities, can be estimated using typical experimental settings with $^{87}$Rb atoms~\cite{Lin2011, Wu2016,Sun2018} with $\omega_{z} =2\pi\times 150\,$Hz,  $k_L\approx\,3\mu{\rm m}^{-1}$ and $k_R\approx 6\,\mu{\rm m}^{-1}$, yielding $E_L\approx3.5\times10^{-31}$J. Then $\gamma\approx 2$, $N_0\sim 200$. The Rabi frequencies $\Omega_{0,1}$  explored below can vary over a range of dozen recoil energies. 

\rev{We are looking for stationary solutions $\bPsi(\br,t)=e^{-i\mu t}\bpsi(\br)$, where $\mu$ is the chemical potential, and $\bpsi=(\psi_1,\psi_2)^{T}$ solve the stationary GPE
\begin{align}
    \mu\bpsi=H\bpsi+g(\bpsi^{\dagger}\bpsi)\bpsi.
\end{align}
}
The phenomena described in this work are based on existence of an extremely flat band in the spectrum $\mu_\nu(\bm{k})$ of $H$: $H\bvarphi_{\nu\bm{k}}=\mu_\nu(\bm{k})\bvarphi_{\nu \bm{k}}$, where $\bvarphi_{\nu \bm{k}}$ is the two-component linear Bloch state, $\nu=1,2...$ is the band number, $\bk=(k_x,k_y)$ is the Bloch wavevector in the first Brillouin zone (BZ) $k_{x,y}\in [-1,1)$. The flatness of the band can be characterized by the bandwidth~\cite{Wang2023,Wang2024a}: 
\begin{align}
    \Delta=\max_\bk\mu_0(\bk)-\min_\bk\mu_0(\bk).
\end{align}

\begin{figure}[t]
	\includegraphics[width=\columnwidth]{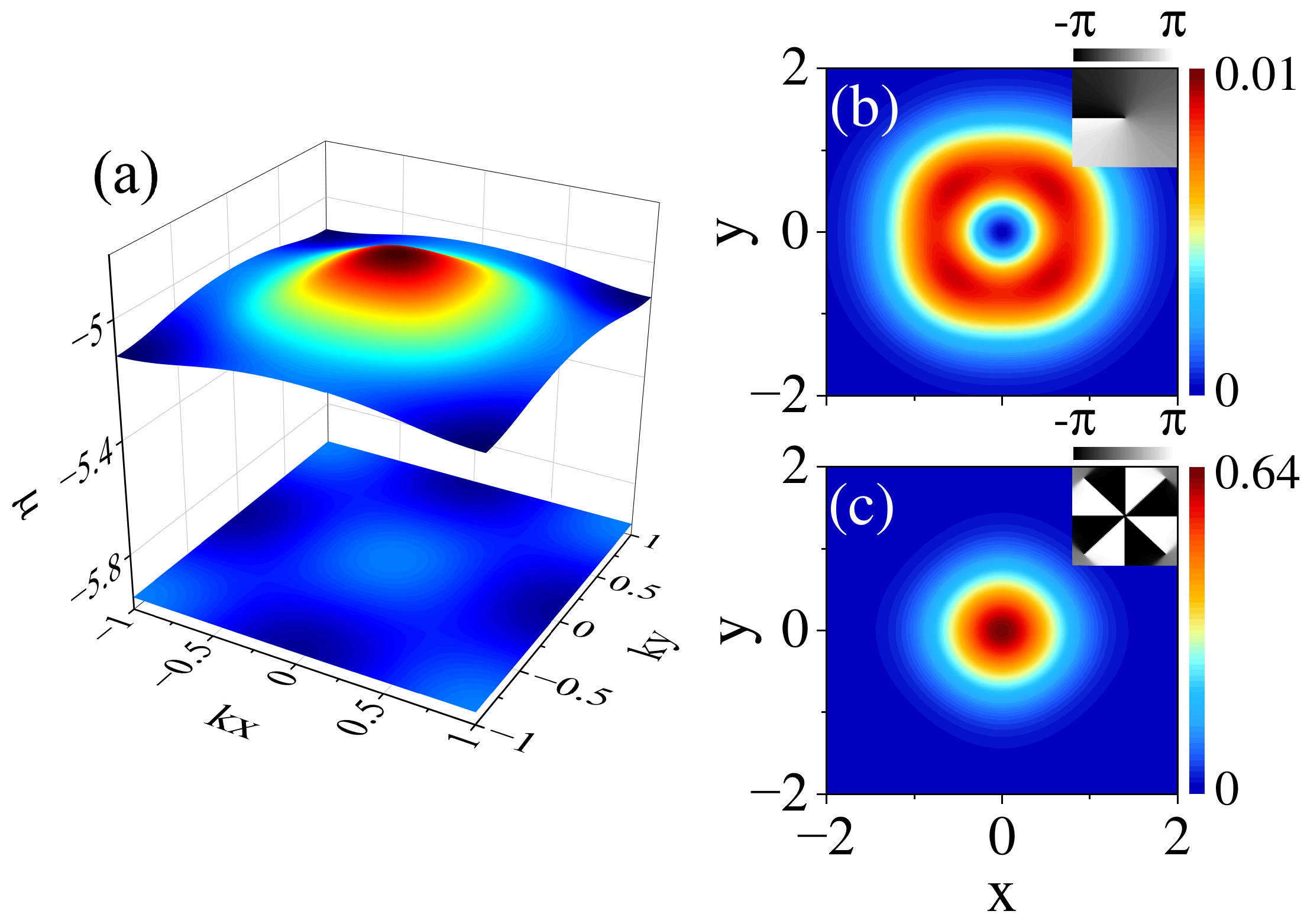}
	\caption{(a) Two lowest bands of the Hamiltonian (\ref{H}). The lowest band at $\max_\bk\mu_0(\bk)=-5.9563$ (corresponding to $\Gamma$ point of the Brillouin zone) and $\min_\bk\mu_0(\bk)=-5.958$ (corresponding to X point of the Brillouin zone) has flatness $\Delta= 1.7 \times 10^{-3}$ (curvature of its landscape is indistinguishable on the scale of the figure). (b,c) The first, $\psi_1$, and second, $\psi_2$ components of the single half-vortex soliton in the respective ZL at  $g=1$ corresponding to $\mu=-5.8563$: the corresponding distributions of the phase are shown in the insets. In all panels $\gamma=2$, $\Omega_0=5.142$, and $\Omega_1=1$.
    }
        \label{fig:one}
\end{figure}

The Hamiltonian $H$ is separable, which means that the search for system parameters ensuring a flat lowest band reduces to solving two 1D problems corresponding to the $x$ and $y$ directions. In Ref.~\cite{Wang2023}, it was shown that in 1D a flat band can be obtained by varying the constant Rabi frequency $\Omega_0$, which thus serves as a control parameter. It should be noted that the Hamiltonian considered in Ref.~\cite{Lobanov2014} did not have a constant component of the Rabi frequency ($\Omega_0=0$), and the spectrum was gapless, which inhibited the existence of a flat band in that model. In Fig.~\ref{fig:one}(a) the flat lowest band with $\Delta=1.7\times10^{-3}$, and the second band, which is not flat, are illustrated.

\section{Families of single half-vortex solitons}
\label{single}

In addition to discrete translational symmetry, the Hamiltonian (\ref{H}) has a Klein four-group $K_4$ of symmetries \{$1, \halpha_1, \halpha_2, \halpha_3$\} where
$\halpha_1=-\mathcal{P}\sigma_z$, $\halpha_2=\mathcal{P}_y\mathcal{K}$, $\halpha_3=-\mathcal{P}_x\mathcal{K}\sigma_z$, $\mathcal{P}_{x,y}$ are the inversion operators along the respective directions, $\mathcal{P}=\mathcal{P}_{x}\mathcal{P}_{y}$, and $\mathcal{K}$ is the complex conjugation operator. Hamiltonian (\ref{H}) possesses a two-dimensional rotational symmetry corresponding to the cyclic group $\mathbb{Z}_4$, whose generator is given by   $\hat r=\text{diag}(i,1)\mathcal{R}(-\pi/2)$, where $\mathcal{R}(\theta)$ is the operator of 2D rotation through the angle $\theta$. Note that this symmetry is broken for the gauge field considered in Ref.~\cite{Lobanov2014}. This additional point symmetry subgroup, $\{1,\hat r, \hat r^2,\hat r^3\}$, leads to several notable differences of the half-vortexes considered here, compared to previously considered ones.  Noting that $\hat r^{2}=\halpha_1$ and $\hat r^4=1$ one obtains the sub-group of local symmetries of the Hamiltonian (\ref{H}),
\begin{align}
    \mathfrak{D}_{\rm loc}=\left\{1,\halpha_1,\halpha_2,\halpha_3,\hat r,\hat r\halpha_1,\hat r\halpha_2, \hat r\halpha_3\right\},
\end{align}
which is isomorphic to the Dihedral group with order 8. Note the similarity of $\mathfrak{D}_{\rm loc}$ with the local symmetry group of BEC Hamiltonian with a helicoidal SOC~\cite{Kartashov2020a}.  

An $\halpha_1$-symmetric solution $\psi_{j}=\left[n_{j}(\br)\right]^{1/2}e^{i\theta_{j}(\br)}$ ($j=1,2$) with the atomic density $n_{j}(\br)$ has phases obeying the following relations $\theta_{1}(\br)=\theta_{1}(-\br)+\pi$ and $\theta_{2}(\br)=\theta_{2}(-\br)$. This implies that at $\br=0$ either a phase singularity or a phase jump should be observed. Both possibilities are indeed observed the model of Ref.~\cite{Lobanov2014}, that does not possess $\mathbb{Z}_4$ symmetry group. In contrast, in our case such symmetry implies that $\theta_{1}(\br)=\mathcal{R}(-\pi/2)\theta_{1}(\br)+\pi/2$ and $\theta_{2}(\br)=\mathcal{R}(-\pi/2)\theta_{2}(\br)$. Thus, the highly symmetric fundamental one-hump gap solitons must carry a central phase singularity at $\br=0$ (and cannot manifest a phase jump) in only the first component, while the second component can not carry vorticity as shown in Fig.~\ref{fig:one} (b,c).  
The numerical results indicate that the winding number $\ell$ at $\br=0$ of the first component is $\ell=-1$, while for the center of the second component $\ell=0$. The ground state is nondegenerate with the specific $\mathbb{Z}_4$ symmetry determining the sign of the winding number. We also observe that the first component bearing vorticity has much smaller intensity than the second one, which is explained by a relatively large constant component of the Rabi frequency.

\begin{figure}[t]
	\includegraphics[width=\columnwidth]{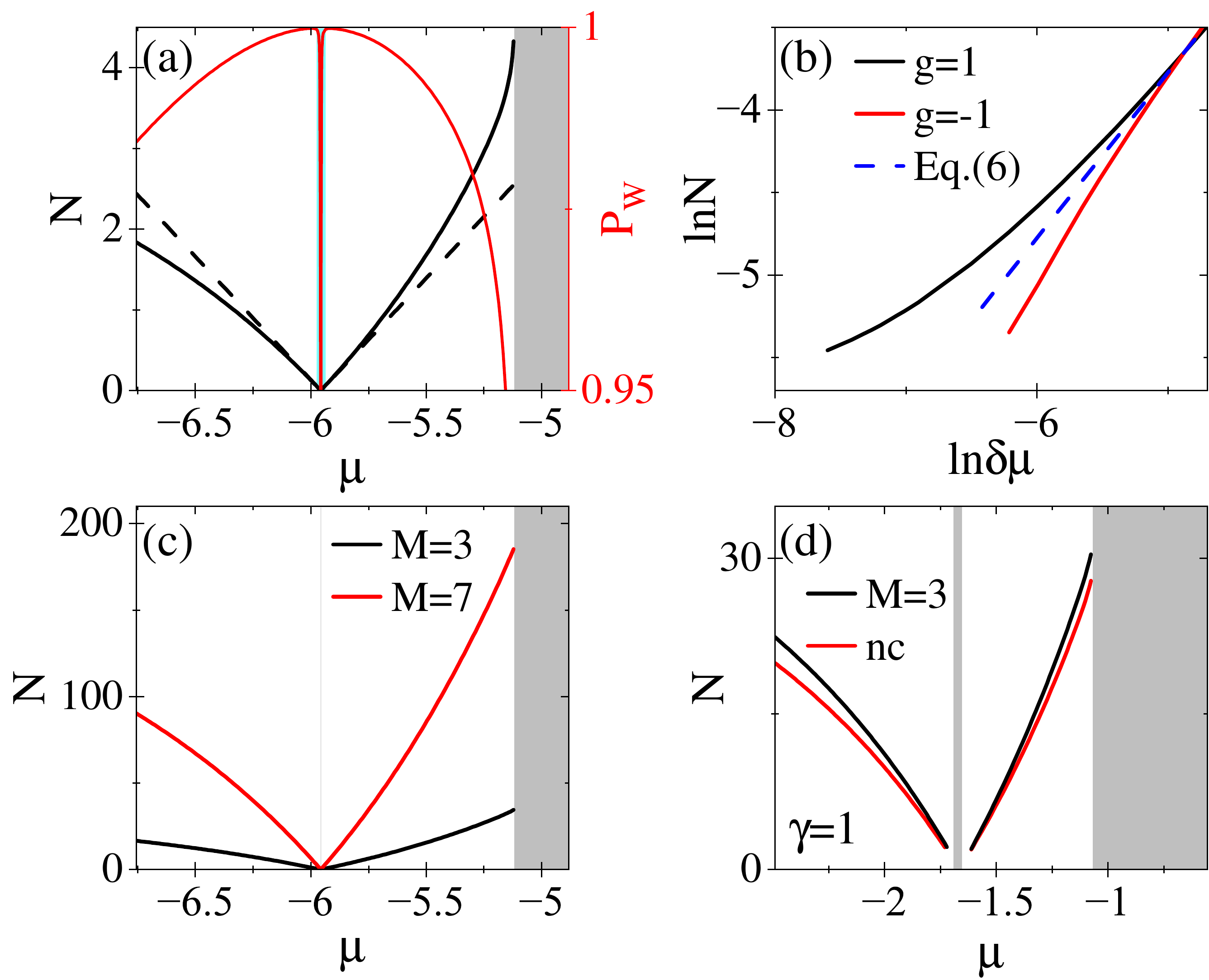}
	\caption{Families of single half-vortex solitons (a,b) and half-vortex lattices (c) for $g=-1$ in the semi-infinite gap $\mu<\min_\bk\mu_0(\bk)\approx-5.958$ and for $g=1$ in the first finite gap 
    $-5.1178>\mu>\max_\bk\mu_0(\bk)\approx-5.9563$. In (a) and (c), the band with flatness $\Delta=1.7\times10^{-3}$ is indistinguishable on the scale of the figure and is represented by a tiny gray stripe. In (a), black solid lines are for one-hump soliton families
    while red lines show $P_W$. The dashed lines represent (\ref{N_mu}) with $\chi=0.3277$. (b) The ln-ln plot zooms the vicinity of the flat band (the  position of the linear band is located at $-\infty$ of the abscissa).  (c) Families of $3\times3$ (black lines) and $7\times7$ (red lines) half-vortex lattices. Other parameters of (a-c) are the same as Fig.~\ref{fig:one}. (d) Families of $3\times3$ half-vortex lattices (black lines with label M=3) and $3\times3$ half-vortex lattices without the cental vortex  (red lines with label nc) are shown in the semi-finite gap [$\mu<\min_\bk\mu_0(\bk)\approx-1.6913$ (X point of the BZ)] and the first finite gap [$-1.0763>\mu>\max_\bk\mu_0(\bk)\approx-1.6528$ ($\Gamma$ point of the BZ)] for  $\gamma=1$, $\Omega_0=0.998$, and $\Omega_1=1$. The band flatness heres is $\Delta=0.038$. 
    }
        \label{fig:two}
\end{figure}

In the proximity of the flat band, half-vortex solitons remain localized on the scale of one lattice period, having shapes well approximated by linear WFs $\bw(\br)$ (see the discussion in~\cite{Wang2024a,Shen2024}). To obtain WFs numerically, one can follow the procedure described in \cite{Marzari2012} with the smooth gauge. In Fig.~\ref{fig:two}(a) we plot the families of half-vortex solitons (black lines) and their projections on the corresponding WFs (i.e., WFs centered at the same lattice site) $P_W=\int\bw_0^{\dagger}\bpsi d\br/\sqrt{N} $ for both attractive and repulsive BECs (red lines). We find that within a relatively broad range of chemical potentials around the flat band, the projection coefficients $P_W$ remain close to unity, except in the immediate vicinity of the flat band, where they exhibit a sharp decrease. In these regions, the half-vortex solitons become significantly broader (as the band is only quasi-flat, rather than perfectly flat), which brings them beyond the limits of reliable numerical accuracy. Consequently, such solutions cannot be obtained numerically.

In the vicinity of the flat band (but beyond the mentioned tiny domain where solitons become wide), the one-soliton family parametrized by $N(\mu)$ is a {\em linear} function of the chemical potential:
\begin{align}
\label{N_mu}
    N(\mu)=\delta\mu/\chi,
\end{align}
 where 
\begin{align}
 \delta\mu=
 \begin{cases}
     \mu-\max_\bk\mu_0(\bk) & g=1
     \\
     \min_\bk\mu_0(\bk)-\mu & g=-1
 \end{cases}   
\end{align}
(note that $\delta\mu>0$) and 
\begin{align}
    \chi=\int(\bw^{\dagger}\bw)^2d\br
\end{align}
is the inverse participation ratio of the normalized WFs.
The dependence  $N(\mu)$ is illustrated in Figs.~\ref{fig:two}(a) by dashed lines.  In Fig.~\ref{fig:two}(b),
we show zoom of the families in the domain of validity of the approximation (\ref{N_mu}) in the ln-ln plot.
 
The details of the derivation of the relation (\ref{N_mu}) are presented in Ref.~\cite{Wang2024a,Shen2024} and remain valid for our system. 
Here (\ref{N_mu}) is verified by obtaining a direct numerical solution of Eq.~(\ref{GPE}) using both Newton relaxation and difference methods~\cite{Yang2010}. The single half-vortex soliton families for the attractive and repulsive interaction are shown by black solid lines Fig.~\ref{fig:two}(a), while the dashed lines are for show the law ~(\ref{N_mu}). Families of fundamental half-vortex solitons do not bifurcate from the linear flat band, although they closely approach it (the difference is not discernible on the scale of the panel). Therefore, in Fig.~\ref{fig:two}(b), we zoom in on the corresponding regions and present them on ln–ln scales, where the linear band is located at $-\infty$ of the abscissa. One can see that the excitation threshold of fundamental half-vortex solitons is indeed anomalously small, and that the families follow the predicted dependence (\ref{N_mu}), except within a narrow region near the linear band. The most remarkable observation is that apparently there is not significant differences in behavior of the families near the flat band for attractive ($g=-1$) and repulsive ($g=+1$) condensates, what contrasts with the typical behavior in the case of non-flat bands.

In the meantime, the predicted relation (\ref{N_mu}) breaks down when the deviation from the flat band becomes appreciable. Indeed, Fig.~\ref{fig:two}(a) shows a clear divergence between the solid and dashed black lines as $\delta\mu$ increases, particularly in the region near the second band. This behavior can be attributed to the distinct properties of strongly attractive or repulsive condensates, where the influence of the flat band is no longer dominant.

Finally, we observe that the presence of the flat band enhances the stability of the half-vortex solitons. To verify this, we performed a linear stability analysis by \rev{computing the maximum of the imaginary part of the spectral parameter, $\lambda_{\rm max}$, of} the corresponding Bogoliubov–de Gennes equations \rev{[see Appendix~\ref{BdG}]}. We also carried out direct numerical simulations of the long-time dynamics governed by Eq.~(\ref{GPE}).
In the latter case \rev{of real-time propagation}, a weak Gaussian noise with an amplitude of about $5\%$ of the soliton profile was added to the initial condition. Both approaches yielded consistent results, showing in particular that single half-vortex solitons, whose families are presented in Figs.~\ref{fig:two}(a) and \ref{fig:two}(b), remain stable \rev{with $\lambda_{\rm max}=0$}  throughout the entire parameter range where the relation (\ref{N_mu}) is valid.

\section{Half-vortex lattices}
\label{lattice1}

The strong localization of single half-vortex solitons supported by the flat band suggests the possible existence of {\em half-vortex lattices}. These higher-order solitary structures can be regarded as composites of fundamental half-vortex solitons and respect the full symmetries $K_4$ and $\mathbb{Z}_4$.

\begin{figure}[t]
	\includegraphics[width=\columnwidth]{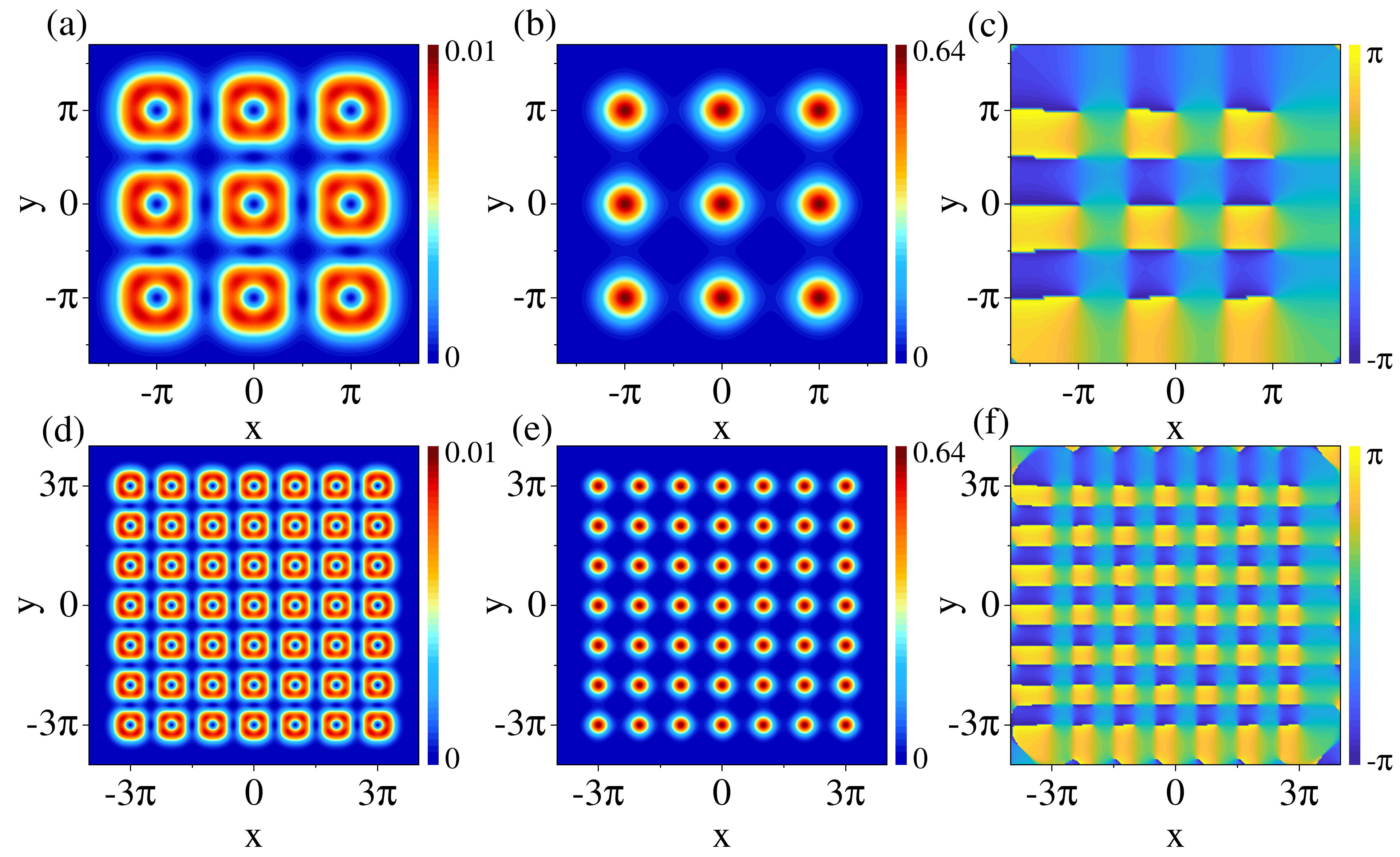}
	\caption{Density distribution of the (a,b) $3\times3$ and (e,f) $7\times7$ half-vortex lattices. The corresponding phase diagrams of the first components are shown in panels (c) and (f), respectively. Other parameters are the same as the Fig.~\ref{fig:one}.}
        \label{fig:three}
\end{figure}

Unlike the single half-vortex soliton and the Bloch states (with infinite array of embedded vortexes)~\cite{Sakaguchi2013,Sakaguchi2016}, the half-vortex lattices considered here are of an arbitrary spatial size. In Fig.~\ref{fig:three}, we present examples of half-vortex lattices of size $M\times M$, where $M$ denotes the number of single half-vortices arranged along each spatial direction. \rev{The initial states for such arrays of half-vortices are constructed by superposing single half-vortex solitons located in different lattice cells.} 

In the left two column of Fig.~\ref{fig:three}, we illustrate the $3\times3$ (upper row) and $7\times 7$ (bottom row) vortex lattices in the real space while the respective phase diagrams are shown in the rightmost column. An important property, which can be seen in the phase distributions, is that while the lattices look like a "combination" of half-vortex solitons ($\ell=-1$), appeared in the density distributions in the leftmost and central panels, phase singularities with winding numbers $\ell=+1$ in the first component appear between two adjacent half-vortex solitons. 

Thus, the lattices are not a simple superposition of individual half-vortex solitons, but rather a single, self-consistent multi-hump nonlinear solution. Additionally, the complex structure of the lattices manifests itself in slightly deformed shaped of "single" half-vortex solitons composing the lattice.
Thus, in panels (c) and (f) of Fig.~\ref{fig:three}, we obtain $M^2+(M-1)^2$ phase singularities with $\ell=-1$ and $2M(M-1)$ singularities with $\ell=+1$. Thus, the total number of phase singularities is $(2M-1)^2$, while total vorticity of the lattice solution remains $\ell=-1$.

In Fig.~\ref{fig:two}(c), families of half-vortex lattices with $M=3$ (black lines) and $M=7$ (red lines) are exhibited. Numerical results show that $N_{M=3}/9\sim N_{M=7}/49$, i.e., $N_M/M^2$ is weakly dependent on $M$. This can be interpreted as almost equal contribution of "single" half-vortex making up the lattice independently on their size, what is explained by  negligible overlaps between nearest half-vortex solitons due to their localization on the one-cell area. Meanwhile, we found that the half-vortex lattices shown in Fig.~\ref{fig:two}(c) share all symmetries with the Hamiltonian and are stable only in the case of attractive interactions ($g=-1$) \rev{[see also Appendix.~(\ref{BdG})]}.

\begin{figure}[t]
	\includegraphics[width=\columnwidth]{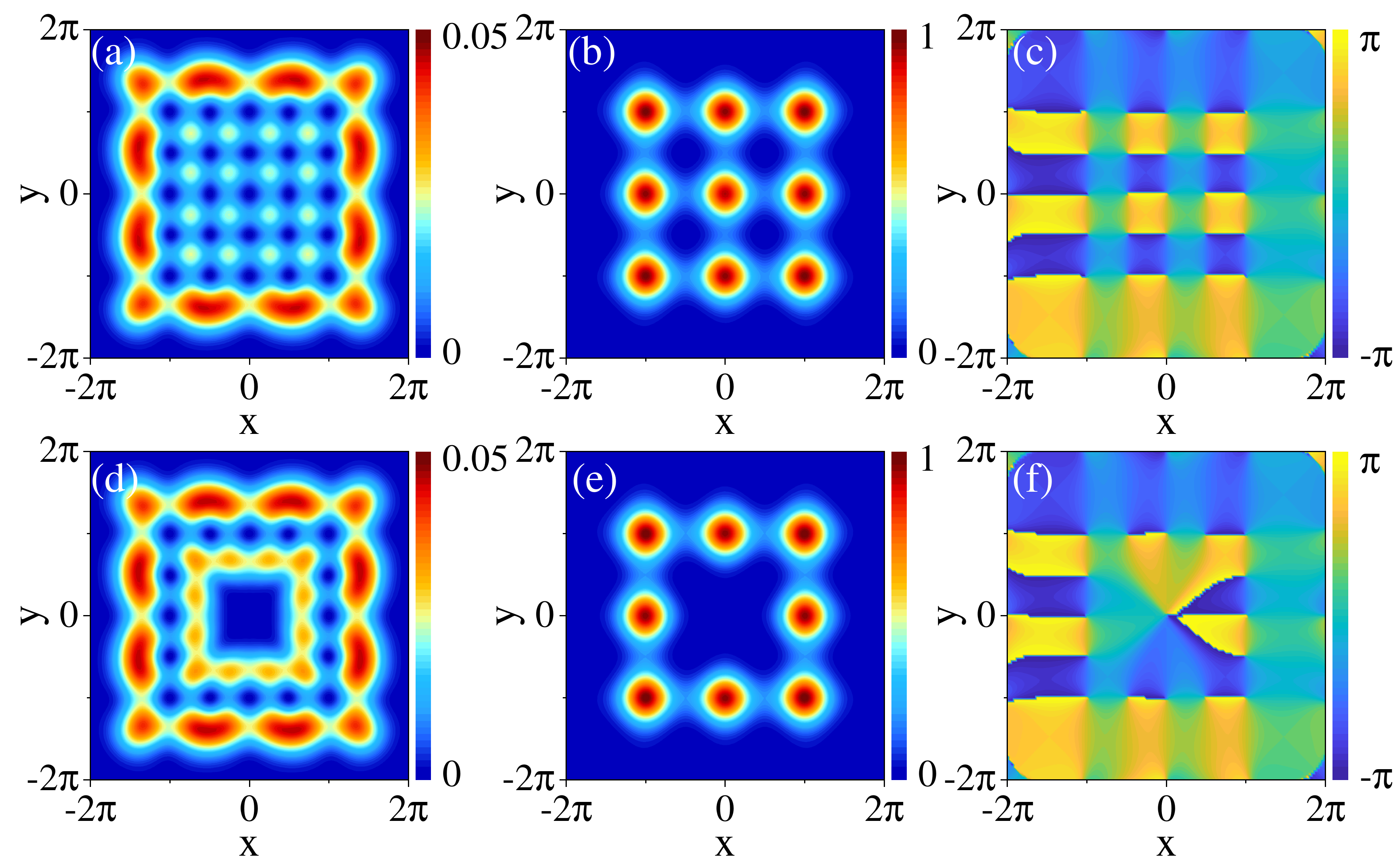}
	\caption{Density distribution of (a,b) the half-vortex lattices and (d,e) half-vortex lattices without the central half-vortex soliton. The corresponding phase diagram is shown in panels (c) and (f), respectively. Here we set $g=1$, $\gamma=1$, $\mu=-1.19$ and other parameters are the same as Fig.~\ref{fig:two}(d).}
        \label{fig:four}
\end{figure} 

Recall that relatively large values of $\gamma$ and $\Omega_0$ must be chosen in order to ensure that the lowest band remains as flat as possible. We now consider the behavior of half-vortex lattices when the band lacks extreme flatness, although still sufficiently flat. To this end we explore the SOC strength $\gamma=1$, noting that 
$$
\Delta_{\gamma=0}=0.17>\Delta_{\gamma=1}=0.038>\Delta_{\gamma=2}=1.7\times10^{-3}.
$$ 
In  Fig.~\ref{fig:two}(d), we show a whole branch (the black lines) of $M=3$ half-vortex lattices with $\gamma=1$, where the threshold of their excitation can be clearly observed. Such modes no longer exist once $N\lesssim2$; thus, a quasi-flat band (with high flatness) is essential to quasi-linear excitations of half-vortex lattices.

\begin{figure*}[t]
	\includegraphics[width=\textwidth]{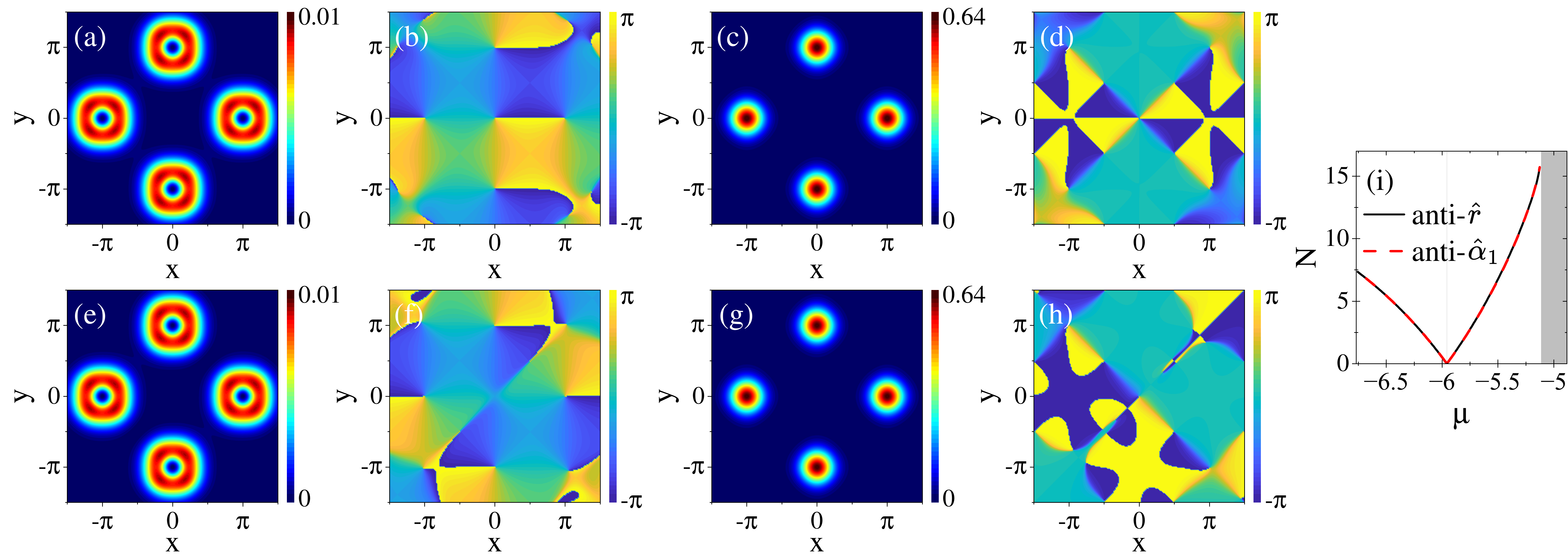}
	\caption{Density distribution, phase diagram, and branches of two types of super-half-vortexes. For (a,c), density distribution of anti-$\hat{r}$ symmetric super-half-vortexes with relative phase $\pi$. Corresponding phase diagrams of two components are shown in panels (b) and (d), respectively. Similarly, for (e-h), density distribution and corresponding phase diagram of anti-$\halpha_1$ symmetric super-half-vortexes are exhibited, respectively. (i) Families of fully symmetric (black solid lines) and anti-$\halpha_1$ symmetric half-vortex lattices are shown in the semi-finite gap ($g=-1$) and the first gap ($g=1$). Here, the vertical gray lines indicate the bands. The parameters are the same as the Fig.~\ref{fig:one}.}
        \label{fig:five}
\end{figure*}

Typical density distributions of the components and phases of the first component are exhibited in the upper row of Fig.~\ref{fig:four}.   One observes that the second component (without vorticity) represents clear "bounds" between the neighboring cells [see Fig.~\ref{fig:four}(b)]. The second component bearing vorticity show strongly inhomogeneous density distributions with most atoms being concentrated along the lattice perimeter [see the appearance of a wall encircling the entire vortex lattices in Fig.~\ref{fig:four}(a)]. The total vorticity of the first component remains one ($\ell=-1$), indicating that the system's symmetries still protect those singularities.

Interestingly, the robustness of the half-vortex solitons which are supported by a flat band and are described by the mutually orthogonal Wannier states modulated by the nonlinearity  one can construct new solutions by removing the central half-vortex soliton. Families of such solutions are shown in Fig.~\ref{fig:two} (d), and an explicit example is presented in the bottom row of Fig.~\ref{fig:four}. In Fig.~\ref{fig:four}(d) one observes an additional "inner wall" in the density distribution of the second component.  As before, the total vorticity of the entire half-vortex lattice remains $\ell=-1$, while a new vortex is created at the center. Solutions of this kind, which do not include the central half-vortex soliton, have been found to be stable for attractive interactions and unstable for repulsive ones.

\section{Super-half-vortexes}
\label{super}

Drawing a parallel to the method of utilizing single half-vortexes for creating lattices and arrays, as previously mentioned, we propose that a vortex can likewise be formed by merging half-vortex solitons. Such a composite state will be referred to as a "super-half-vortex soliton".
Figure~\ref{fig:five} illustrates an example comprising four vortexes arranged in two different ways. 

In the upper row of Fig.~\ref{fig:five}, four vortexes are centered in the vertices of a square bearing relative phases $\pi$ with respect to their neighbors. This solution still obeys the symmetries of the Klein-four group $K_4$, but becomes {\em anti}-$\hat{r}$ symmetric.  Thus, a vortex with $\ell=1$ is created at the center due to $\theta_1(\br)=\mathcal{R}(-\pi/2)\theta_1(\br)-\pi/2$ as shown in Fig.~\ref{fig:five}(b). In contrast, at the soliton component, the density also vanishes at $\br=0$ while satisfying $\theta_2(\br)=\mathcal{R}(-\pi/2)\theta_2(\br)-\pi$ [see Fig.~\ref{fig:five}(d)], i.e., a giant half-vortex with the winding number $\ell=-2$ of the first component is created. Thus, such four-hump solutions are termed anti-$\hat{r}$ symmetric super-half-vortex solitons and represent {\em nonlinear symmetry breaking} (we note that symmetry breaking of 2D solitons in BEC with a helicoidal SOC was described in~\cite{Kartashov2020a}).

The situation can be quite different if one changes the relative phases. In the bottom row of Fig.~\ref{fig:five}, we show anti-$\halpha_1$ symmetric solution which breaks all local symmetries $\mathfrak{D}_{\rm loc}$. One does not observe any singularity at $\br=0$ of the first component. However, a vortex with $\ell=-1$ is created as expected at the second component. In other words one obtains an anti-$\halpha_1$ symmetric super-half-vortex soliton. By applying the $\halpha_j$ or $\hat r$ one can obtain other branches of the symmetry-broken solutions, although not all being the new ones. For example, we obtained that $\hat r \bpsi$  and $\halpha_2 \bpsi$ represent the same solution.

Anti-$\hat{r}$ symmetric super-half-vortex solitons appear to be stable only in SOC-BECs with purely repulsive interactions, whereas anti-$\hat{\alpha}_1$ symmetric super-half-vortex solitons are found to be always unstable \rev{[see also Appendix.~(\ref{BdG})]}. In Fig.~\ref{fig:five}(i), we show the families of both types of super-half-vortex solitons, which are practically indistinguishable, differing only slightly.

\section{Conclusion}
\label{conclusion}
In this work, within the mean-field approximation, we have described families of half-vortex lattices that emerge in two-dimensional spin–orbit–coupled Bose–Einstein condensates loaded into Zeeman lattices whose linear spectrum exhibits a \rev{quasi}-flat lowest band. In such systems, excitations of half-vortex solitons occur with a negligible (although non-zero) threshold number of atoms. Respectively, such solitons are well described by modulated Wannier functions. The local symmetries, isomorphic to a dihedral group of order 8, of the system impose a constraint on the winding number of one of the components. Strong confinement of solitons within practically a single lattice cell allows for the construction of lattices of solitons of a wide range of geometrical forms. In particular, a super-half-vortex soliton, which can be viewed as properly arranged interacting single-half-vortex solitons, has been found. We also obtained numerically the nonlinear symmetry breaking, i.e., half-vortex solitons that do not obey symmetries of the governing linear Hamiltonian. The use of the identified symmetries allows for further systematic construction of multi-vortex solutions breaking linear symmetries.

\section*{Acknowledgments}

The work was supported by the Fundação para a Ciência e Tecnologia under the projects 2023.13176.PEX (DOI https://doi.org/10.54499/2023.13176.PEX) and by FCT – Fundação para a Ciência e a Tecnologia, I.P., through national funds, under the Unit CFTC - Centro de Física Teórica e Computacional, reference UID/00618/2023, financing period 2025-2029.
This work was also supported by the National Natural Science Foundation of China (NSFC) under Grants No.
12374247 and No. 11974235, as well as by the Shanghai
Municipal Science and Technology Major Project (Grant
No. 2019SHZDZX01-ZX04).

\appendix

\section{Linear stability analysis}
\label{BdG}
 
\rev{Here we present the details of the linear stability analysis. As is customary, we seek the perturbed solution of Eq.~(\ref{GPE}) in the form $e^{-i\mu t}(\bpsi+\delta\bpsi)$ where $\delta\bpsi=(\delta\psi_1,\delta\psi_2)^T$, 
\begin{align}
    \delta\psi_j=u_{j}\exp\left(-i\lambda t\right)+v_{j}^{*}\exp\left(i\lambda^{*}t\right),
\end{align}
where $j=1,2$, $u_{j}$ and $v_{j}$ are perturbation amplitudes, and $\lambda$ is the spectral parameter. Keeping only linear (with respect to $u_j$ and $v_j$) terms, we obtain the  Bogoliubov-de Gennes (BdG) equations:
\begin{align}
    \mathcal{H}_{\text{BdG}}\bm{U}=\lambda\bm{U}.
\end{align}
Here $\bm{U}=(u_1,u_2,v_1,v_2)^T$, and the BdG Hamiltonian,
\begin{align}
    \mathcal{H}_{\text{BdG}}=\left(\begin{array}{cc}
        H-\mu+\mathcal{A} & \mathcal{B} \\
        -\mathcal{B}^* & -(H-\mu+\mathcal{A})^* 
    \end{array}\right),
\end{align}
with 
\begin{align*}
    \mathcal{A}&=g\left(\begin{array}{cc}
       2|\psi_1|^2+|\psi_2|^2  & \psi_1\psi_2^* \\
       \psi_1^*\psi_2  & 2|\psi_2|^2+|\psi_1|^2
    \end{array}\right),\\
    \mathcal{B}&=g\left(\begin{array}{cc}
       \psi_1^2  & \psi_1\psi_2 \\
       \psi_1\psi_2  & \psi_2^2
    \end{array}\right)
\end{align*}
(the asterisk denotes complex conjugation). The instability of the solutions is characterized by the maximum of imaginary parts of the eigenvalues $\lambda$, i.e., by $\lambda_{\rm max}$. When $\lambda_{\rm max}=0$ is the corresponding nonlinear state is stable.}

\begin{figure}[b]
    \centering
    \includegraphics[width=1\linewidth]{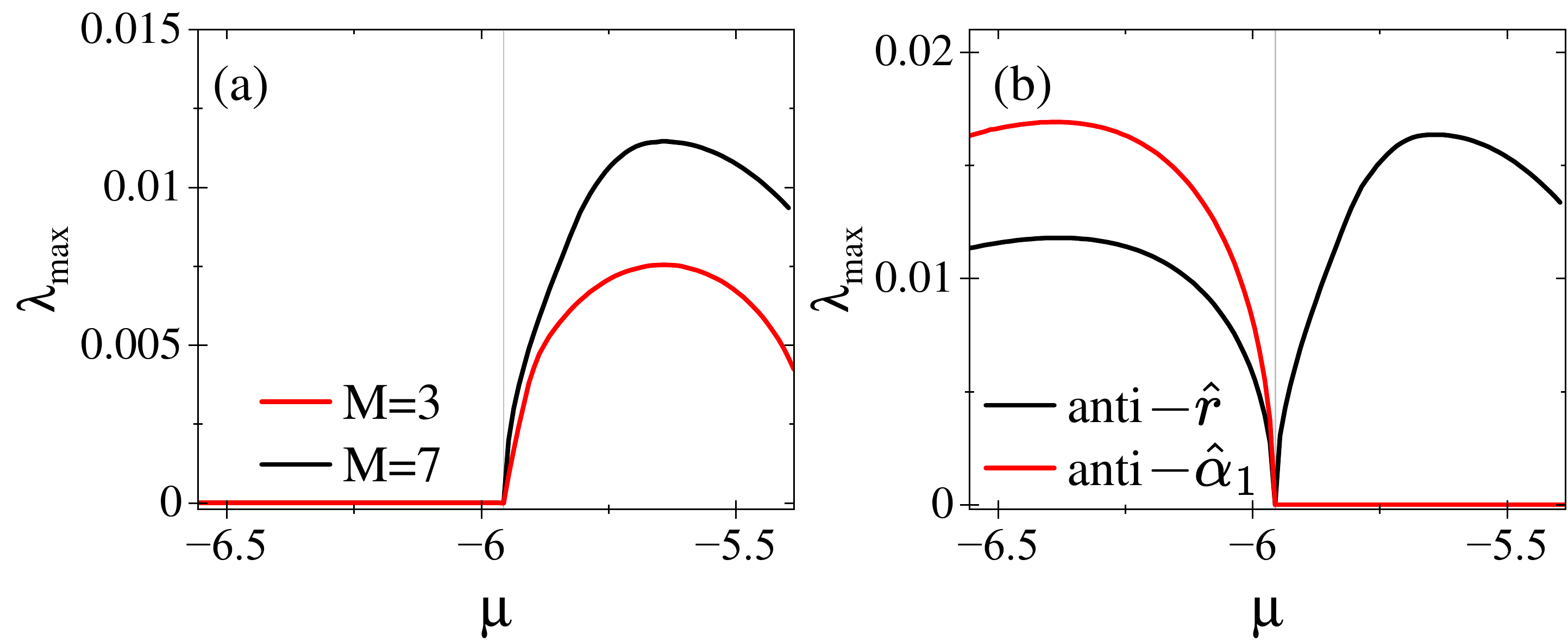}
    \caption{\rev{The stability regions of (a) half-vortex lattices and (b) super-half-vortexes for $g=-1$ in the semi-infinite gap and for $g=1$ in the first gap. The vertical gray line indicates the band. Other parameters are the same as in Figs.~\ref{fig:two}(c) and Fig.~\ref{fig:five}(i), respectively.}}
    \label{fig:s1}
\end{figure}

\rev{Typical results for stability of half-vortex lattices and super-half-vortexes are shown in Fig.~\ref{fig:five}, which are discussed in the main text.}


\printcredits


\begin{thebibliography}{100}

\bibitem{Malomed2024}
Malomed, B.A., 2024. Multidimensional soliton systems. Advances in Physics: X 9, 2301592.

\bibitem{Mihalache2024}
Mihalache, D., 2024. Localized structures in optical media and Bose-Einstein condensates: An overview of recent theoretical and experimental results. Rom. Rep. Phys. 76, 402.

\bibitem{Han2012} 
Han, W., Zhang, S., Jin, J., Liu, W.M., 2012. Half-vortex sheets and domain-wall trains of rotating two-component Bose-Einstein condensates in spin-dependent optical lattices, Phys. Rev. A 85, 043626.

\bibitem{Zamora2012} 
Zamora-Zamora, R., Lozada-Hidalgo, M., Caballero-Ben\'itez, S.F.,  Romero-Roch\'in, V., 2012. Vortices on demand in multicomponent Bose-Einstein condensates, Phys. Rev. A 86, 053624.

\bibitem{Kasamatsu2013}  
Kasamatsu, K., Takeuchi, H., Tsubota, M., Nitta, M., 2013. Wall-vortex composite solitons in two-component Bose-Einstein condensates, Phys. Rev. A 88, 013620.
 
\bibitem{Wen2013} 
Wen, L., Qiao, Y., Xu, Y., Mao, L., 2013. Structure of two-component Bose-Einstein condensates with respective vortex-antivortex superposition states, Phys. Rev. A 87, 033604.

\bibitem{Zhang2014}
Zhang, X.F., Dua, Z.J., Tanb, R.B.,
Donga, R.F., Changa, H., Zhang,  S.G., 2014. Vortices in a rotating two-component Bose–Einstein condensate with tunable interactions and harmonic potential, Annals of Physics 346, 154-163 (2014). 

\bibitem{Xu2005} 
Xu, Z., Kartashov, Y.V., Crasovan, L.C., Mihalache, D., Torner, L., 2005. Multicolor vortex solitons in two-dimensional photonic lattices,
Phys. Rev. E 71, 016616.

\bibitem{Wang2008} 
Wang, J., Yang, J., 2008. Families of vortex solitons in periodic media, Phys. Rev. A 77, 033834.

\bibitem{Law2010} 
Law, K.J.H., Kevrekidis, P.G., Tuckerman, L.S., 2010. Stable Vortex–Bright-Soliton Structures in Two-Component Bose-Einstein Condensates, Phys. Rev. Lett. 105, 160405.

\bibitem{Stockhofe2011} 
Stockhofe, J., Kevrekidis, P.G., Frantzeskakis, D.J., Schmelcher, P., 2011.
Dark–bright ring solitons in Bose–Einstein condensates J. Phys. B: At. Mol. Opt. Phys. 44, 191003. 

\bibitem{Pola2012} 
Pola, M.,  Stockhofe, J., Schmelcher, P., Kevrekidis, P.G., 2012. Vortex–bright-soliton dipoles: Bifurcations, symmetry breaking, and soliton tunneling in a vortex-induced double well, Phys. Rev. A  86, 053601.

\bibitem{Sanjay2025}
\rev{Sanjay S., Saravana V.S., Malomed, B.A., 2025, Vortex droplets and lattice patterns in two-dimensional traps: A photonic spin–orbit-coupling perspective. Chaos, Solitons \& Fractals, 197, 116441.}

\bibitem{Deekshita2025}
\rev{Deekshita, S., Sanjay, S., Veni, S.S., Tabi, C.B., Kofane, T.C., 2025. Synergistic Effects of Spin-Orbit Coupling and Intercomponent Interactions in Two-Component (2+1) D Photonic Fields. Chaos, Solitons \& Fractals, 199, 116806.}

\bibitem{Lin2011}
Lin, Y.J., Jim\'enez-Garc\'i, K., Spielman, I.B., 2011. Spin–orbit-coupled Bose–Einstein condensates. Nature 471, 83–8.

\bibitem{Wu2016}
Wu, Z., Zhang, L., Sun, W., Xu, X.T., Wang, B.Z., Ji, S.C., Deng, Y., Chen, S., Liu, X.J., Pan, J.W., 2016. Realization of two-dimensional spin-orbit coupling for Bose-Einstein condensates. Science 354, 83–88. 

\bibitem{Sun2018}
Sun, W., Wang, B.Z., Xu, X.T., Yi, C.R., Zhang, L., Wu, Z., Deng, Y., Liu, X.J., Chen, S., Pan, J.W., 2018. Highly controllable and robust 2D spin-orbit coupling for quantum gases. Phys. Rev. Lett. 121, 150401.

\bibitem{Kartashov2019}
Kartashov, Y.V., Astrakharchik, G.E., Malomed, B.A., Torner, L., 2019. Frontiers in multidimensional self-trapping of nonlinear fields and matter. Nature Reviews Physics 1, 185–197.

\bibitem{Malomed2019}
Malomed, B.A., 2019. Vortex solitons: Old results and new perspectives. Physica D: Nonlinear Phenomena, 399, 108-137.

\bibitem{Sakaguchi2014} 
Sakaguchi, H., Malomed, B.A., 2014. Discrete and continuum composite solitons in Bose-Einstein condensates with the Rashba spin-orbit coupling in one and two dimensions. Phys. Rev. E 90, 062922.

\bibitem{Sakaguchi2016}
Sakaguchi, H., Sherman, E.Y., Malomed, B.A., 2016. Vortex solitons in two-dimensional spin-orbit coupled Bose-Einstein condensates: effects of the Rashba-Dresselhaus coupling and the Zeeman splitting. Phys. Rev. E 94, 032202. 

\bibitem{Lobanov2014}
Lobanov, V.E., Kartashov, Y.V., Konotop, V.V., 2014. Fundamental, multipole, and half-vortex gap solitons in spin-orbit coupled Bose-Einstein condensates. Phys. Rev. Lett. 112, 180403.

\bibitem{Liao2017} Liao, B.; Li, S.; Huang, C.; Luo, Z.; Pang, W.; Tan, H.; Malomed, B.A.; Li, Y. Anisotropic semivortices in
dipolar spinor condensates controlled by Zeeman splitting. Phy. Rev. A 96, 043613 (2017).

\bibitem{Liu2018} Liu, S.; Liao, B.; Kong, J.; Chen, P.; Lü, J.; Li, Y.; Huang, C.; Li, Y. Anisotropic Semi Vortices in Spinor 
Dipolar Bose Einstein Condensates Induced by Mixture of Rashba Dresselhaus Coupling. J. Phys. Soc. Jpn. 87, 094005 (2018).


\bibitem{Kartashov2020a} 
Kartashov, Y.V., Sherman E.Ya., Malomed, B.A., Konotop V.V., 2020, Stable two-dimensional soliton complexes in Bose–Einstein condensates with helicoidal spin–orbit coupling. New J. Phys. 22, 103014.

\bibitem{Wang2024a}
Wang, C., Zhang, Y., Konotop, V.V., 2024. Transfer of solitons and half-vortex solitons via adiabatic passage. Phys. Rev. Research 6, 043089.


\bibitem{Berge1998}
Berg\'e, L., 1998. Wave collapse in physics: principles and applications to light and plasma waves. Physics reports 303, 259–37.

\bibitem{Fibich2015} Fibich, G. 2015 The Nonlinear Schr\"odinger Equation: Singular Solutions and Optical Collapse (Springer, Heidelberg).

\bibitem{Fu2020} 
Fu, Q., Wang, P., Huang, C., Kartashov, Y.V., Torner, L., Konotop, V.V. Ye, F., 2020. Optical soliton formation controlled by angle twisting in photonic moir\'e lattices, Nature Photonics 14, 663.

\bibitem{Ivanov2023} 
Ivanov, S.K., Konotop, V.V., Kartashov, Y.V., Torner, L., 2023. Vortex solitons in moir\'e optical lattices, Optics Letters 48, 3797-3800.

\bibitem{Wang2024b} 
Wang, P., Fu, Q., Konotop, V.V., Kartashov, Y.V., Ye, F. 2024. Observation of localization of light in linear photonic quasicrystals with diverse rotational symmetries, Nature Photonics, 18, 224-229.

\bibitem{Kartashov2020} 
 Kartashov, Y. V., Torner, L., Modugno, M., Sherman, E. Ya., Malomed, B. A. Konotop V. V. 2020, Multidimensional hybrid Bose-Einstein condensates stabilized by lower-dimensional spin-orbit coupling. Phys. Rev. Research, 2, 013036.

\bibitem{Vicencio2013}
Vicencio, R.A., Johansson, M., 2013. Discrete flat-band solitons in the Kagome lattice. Phys. Rev. A 87, 061803.

\bibitem{Leykam2018}
Leykam, D., Andreanov, A., Flach, S., 2018. Artificial flat band systems: From lattice models to experiment. Advances in physics: X 3, 1473052.


\bibitem{Li2025}
Li, H.T., Ji, T.Z., Yan, R.G., Fan, W.L., Zhang, Z.X., Sun, L., Miao, B.F., Chen, G., Wan, X.G., Ding, H.F., 2025. General method to construct flat bands in two-dimensional lattices. Phys. Rev. Lett. 134, 076402.

\bibitem{Shen2024} 
Shen, S., Zhang, Y., Kartashov, Y.V., Li, Y., Konotop, V.V., 2024. Two-dimensional flat-band solitons in superhoneycomb lattices, Nanophotonics, 13, 4047-4056.

\bibitem{Zhang2013b}
Zhang, Y., Zhang, C., 2013. Bose-Einstein condensates in spin-orbit-coupled optical lattices: flat bands and superfluidity. Phys. Rev. A 87, 023611.

\bibitem{Kartashov2016a}
Kartashov, Y.V., Konotop, V.V., Zezyulin, D.A., Torner, L., 2016. Bloch oscillations in optical and Zeeman lattices in the presence of spin-orbit coupling. Phys. Rev. Lett. 117, 215301. 


\bibitem{Hui2017}
Hui, H., Zhang, Y., Zhang, C., Scarola, V.W., 2017. Superfluidity in the absence of kinetics in spin-orbit-coupled optical lattices. Phys. Rev. A 95, 033603.

\bibitem{Wang2023}
Wang, C., Zhang, Y., Konotop, V.V., 2023. Wannier solitons in spin-orbit-coupled Bose-Einstein condensates in optical lattices with a flat band. Phys. Rev. A 108, 013307.

\bibitem{Tai2024}
Tai, Y., Fan, H., Ma, X., Wei, W., Zhang, H., Tang, M., Li, X., 2024. Generation of arbitrarily structured optical vortex arrays based on the epicycle model. Optics Express 32, 10577–10586.

\bibitem{Bychkov1984} 
Bychkov, Y.A., Rashba, E.I., 1984. Oscillatory effects and the magnetic susceptibility of carriers in inversion layers. The Journal of Physical Chemistry C 17, 6039.

\bibitem{Jimenez2015}
Jim\'{e}nez-Garc\'{i}a, K., LeBlanc, L.J., Williams, R.A., Beeler, M.C., Qu, C., Gong, M., Zhang, C., Spielman, I.B., 2015. Tunable spin-orbit coupling via strong driving in ultracold-atom systems. Phys. Rev. Lett. 114, 125301.

\bibitem{Zhang2013a}
Zhang, Y., Chen, G., Zhang, C., 2013. Tunable spin-orbit coupling and quantum phase transition in a trapped Bose-Einstein condensate. Science Report 3, 1937.

\bibitem{Sakaguchi2013}
Sakaguchi, H., Li, B., 2013. Vortex lattice solutions to the Gross-Pitaevskii equation with spin-orbit coupling in optical lattices. Phys. Rev. A 87, 015602. 

\bibitem{Marzari2012}
Marzari, N., Mostofi, A.A., Yates, J.R., Souza, I., Vanderbilt, D., 2012. Maximally localized Wannier functions: Theory and applications. Rev. Mod. Phys. 84, 1419–1475.

\bibitem{Yang2010}
\rev{Yang, J., 2010. Nonlinear waves in integrable and nonintegrable systems. Society for Industrial and Applied Mathematics.}

\end{thebibliography}
\end{document}